\documentclass[prd,aps,showpacs,preprintnumbers,11pt]{revtex4}
\usepackage{epsfig}
\usepackage{subfigure}
\begin{document}  
\newcommand{\be}{\begin{equation}}\newcommand{\ee}{\end{equation}}
\newcommand{\bea}{\begin{eqnarray}}\newcommand{\eea}{\end{eqnarray}}
\newcommand{\bc}{\begin{center}}\newcommand{\ec}{\end{center}}
\def\no{\nonumber}
\def\eq#1{Eq. (\ref{#1})}\def\eqeq#1#2{Eqs. (\ref{#1}) and  (\ref{#2})}
\def\lsim{\raise0.3ex\hbox{$\;<$\kern-0.75em\raise-1.1ex\hbox{$\sim\;$}}}
\def\gsim{\raise0.3ex\hbox{$\;>$\kern-0.75em\raise-1.1ex\hbox{$\sim\;$}}}
\def\slash#1{\ooalign{\hfil/\hfil\crcr$#1$}}
\def\eff{\mbox{\tiny{eff}}}
\def\order#1{{\mathcal{O}}(#1)}
\def\pppm{B^0\to\pi^+\pi^-}
\def\pzpz{B^0\to\pi^0\pi^0}
\def\pppz{B^0\to\pi^+\pi^0}
\preprint{}
\title{CKM Matrix Elements }
\author{T. N. Pham }
\affiliation{
Centre de Physique Th\'{e}orique, CNRS,
Ecole Polytechnique, 91128 Palaiseau, Cedex, France}

\date{\today}
\begin{abstract}
The current analysis of the Cabibbo-Kobayashi-Maskawa(CKM) quark 
mixing matrix uses the standard parametrisation by 3 mixing angles and the
CP-violating KM phase. However it would be more convenient to express
these mixing angle parameters in terms of the known  CKM matrix elements
like $V_{ud}$, $ V_{us}$, $ V_{ub}$, $V_{cb}$ and the CP-violating
phase $\delta$. The other CKM matrix elements are then expressed in terms
of these known matrix elements instead of the standard mixing angles. 
In this paper, using $V_{ud}$, $ V_{us}$, $ V_{ub}$ and  $V_{cb}$
from  the current global fit, we   show that the measured values 
for $|V_{td}/V_{ts}|$ and $\sin\beta$ imply $\gamma=(68\pm 3)^{\circ}$ 
at which $\sin\beta$ and  $\sin\alpha$ are near its maximum, 
consistent with the global fit.

\end{abstract}
\maketitle

\begin{figure}[ht]
\centering
\includegraphics[height=5.0cm,angle=0]{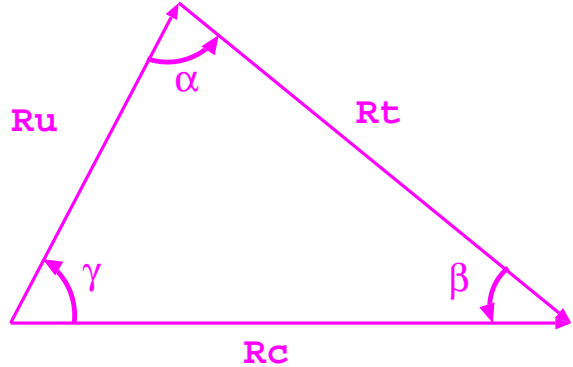}
\caption{The  $(db)$ unitarity  triangle with the sides represent 
$R_{u}=|V_{ud}V_{ub}^{*}/V_{cd}V_{cb}^{*}|$, 
$R_{t}=|V_{td}V_{tb}^{*}/V_{cd}V_{cb}^{*}|$ and 
 $R_{c}= 1$}   
\label{Fig1}
\end{figure}

The current global fit of the CKM matrix elements in the Standard Model 
with imposed  unitarity constraints\cite{Charles,PDG}  seems
to allow a rather  precise determination of some of the less
known CKM matrix elements  like $V_{cb}$ and
$V_{ub}$ . It  would then  be possible
to directly express the mixing angle parameters in terms of the known  
 CKM matrix elements like  $V_{ud}$, $ V_{us}$, 
$ V_{ub}$, $V_{cb}$ and the phase $\delta$,
 rather than using a parametrization for the mixing angles as usually done 
in the current studies  of the CKM matrix elements.  The remaining 
CKM matrix elements    
$V_{cd}$, $ V_{cs}$, $ V_{td}$, $V_{ts}$ and $ V_{tb}$
are  thus completely determined directly
in terms of these known quantities and the CP-violating phase $\delta$,
the  main feature of our approach. Furthermore, the CP-violating 
phase $\delta$  can also be expressed
in terms of the known CKM matrix elements and the angle $\gamma$, $\gamma$
being one of the  angle of the $(db)$ unitarity triangle \cite{PDG,Nir}
as shown in Fig. (\ref{Fig1}).  The other angles 
$\beta$ and $\alpha$ as well as the ratio $|V_{td}/V_{ts}|$ are then  
given as functions of  $\gamma$.  In this paper we shall
present our calculation of  ratio $|V_{td}/V_{ts}|$,  
 $\sin\beta$ and $\sin\alpha$  in terms of $\gamma$ using 
$V_{ud}$, $ V_{us}$, $ V_{ub}$, $V_{cb}$ from the global fit. We
find that  the measured values for $|V_{td}/V_{ts}|$ and 
$\sin\beta$ imply that $\gamma=(68.6\pm 3)^{\circ} $ and 
$\alpha \approx (89\pm 3)^{\circ} $ consistent with the global fit
values for these quantities. 

 We  begin by writing  down 
the Cabibbo-Kobayashi-Maskawa(CKM) \cite{Cabibbo,Kobayashi} quark mixing
matrix as:
\be
V_{\rm CKM} = \pmatrix{V_{ud}& V_{us} & V_{ub} \cr
V_{cd} & V_{cs} & V_{cb} \cr 
V_{td} & V_{ts} & V_{tb}} 
\label{Vckm} 
\ee
To impose unitarity for the CKM matrix, we take the standard 
parametrization as used in \cite{PDG}
which is given as \cite{Chau}:
\be
V_{\rm CKM} = \pmatrix{c_{12}c_{13}& s_{12}c_{13} & s_{13}\exp(-i\delta) \cr
-s_{12}c_{23}-c_{12}s_{23}s_{13}\exp(i\delta) & c_{12}c_{23}- 
s_{12}s_{23}s_{13}\exp(i\delta) & s_{23}c_{13} \cr 
s_{12}s_{23}- c_{12}c_{23}s_{13}\exp(i\delta) & 
-c_{12}s_{23}-s_{12}c_{23}s_{13}\exp(i\delta) & c_{23}c_{13} } 
\label{Chau} 
\ee
where $s_{ij}=\sin(\theta_{ij}$, $c_{ij}=\cos(\theta_{ij}$ and
$\delta$ is the CP-violating KM phase. The angles $\theta_{ij} $
can be chosen to be in the first quadrant, so $s_{ij}$ and  $c_{ij} $
can be taken as positive \cite{PDG}. The  current global fit gives
for the magnitudes of all 9 CKM matrix elements \cite{Charles,PDG}: 
 \be
|V_{\rm CKM}| = \pmatrix{0.97428\pm 0.00015&  0.2253\pm 0.0007 & 0.00347^{+0.00016}_{-0.00012}  \cr
0.2252\pm 0.0007 &0.97345^{+0.00015}_{-0.00016}  & 0.0410^{+0.0011}_{-0.0007} \cr 
0.00862^{+0.00026}_{-0.00020} & 0.0403^{+0.0011}_{-0.0007}  &0.999152^{+0.000030}_{-0.000045} } 
\label{fit}
\ee

Instead of using the current parametrization \cite{Buras,PDG}, 
which is a form of Wolfenstein parametrization \cite{Wolfenstein}
and is unitary to all order in $s_{12}=\lambda$, we shall now express
$s_{ij}$ and $c_{ij}$ in terms of $V_{ud}$, $ V_{us}$, $V_{ub}$ and  $V_{cb}$. 
As these quantities are assumed to be positive, they can be 
obtained directly from the measured
absolute values. For simplicity, we shall put $s_{13}=V_{ub}$
and the matrix element $V_{ub}$ of  the CKM matrix in Eq. (\ref{Vckm}) is now
written with the phase $\delta$ taken out ($V_{ub} \to V_{ub}\exp(-i\delta)$.
We have:
\bea
&& s_{12}=\frac{ V_{us}}{\sqrt{( V_{ud}^{2} + V_{us}^{2})}}, \quad 
c_{12}= \frac{V_{ud}}{\sqrt{( V_{ud}^{2} + V_{us}^{2})}} \nonumber\\
&& s_{13}= V_{ub} ,   \quad \quad c_{13}= \sqrt{(1- V_{ub}^{2})}\nonumber\\
&& s_{23}= \frac{V_{cb}}{ \sqrt{(1- V_{ub}^{2})}}, \quad c_{23}= \frac{\sqrt{(1-
V_{ub}^{2} - V_{cb}^{2})}}{\sqrt{(1- V_{ub}^{2})}} .
\label{sij}
\eea
Since $s_{12}^{2}+ c_{12}^{2}=1 $, we have 
\be
V_{ud}^{2} + V_{us}^{2}= c_{13}^{2} = 1- V_{ub}^{2}
\ee

 The  matrix elements of the CKM matrix in Eq.(\ref{Vckm}) are then
 given by:
\bea
&& V_{us}=\frac{ V_{us}\sqrt{(1- V_{ub}^{2})}}{\sqrt{( V_{ud}^{2} + V_{us}^{2})}},\quad 
V_{ud}= \frac{V_{ud}\sqrt{(1- V_{ub}^{2})}}{\sqrt{( V_{ud}^{2} + V_{us}^{2})}},
 \quad V_{ub}\to V_{ub}\exp(-i\delta)\nonumber\\
&& V_{cd}=-\frac{ V_{us}\sqrt{(1-V_{ub}^{2} - V_{cb}^{2})}}{\sqrt{(
 V_{ud}^{2} + V_{us}^{2})}\sqrt{(1- V_{ub}^{2})}}-\frac{V_{ud}V_{cb}V_{ub}\exp(i\delta)}{\sqrt{( V_{ud}^{2} + V_{us}^{2})}\sqrt{(1- V_{ub}^{2})}} , \nonumber\\
&& V_{cs}= \frac{V_{ud}\sqrt{(1-V_{ub}^{2} - V_{cb}^{2})}}{\sqrt{( V_{ud}^{2} + V_{us}^{2})}\sqrt{(1- V_{ub}^{2})}}-\frac{V_{us}V_{cb}V_{ub}\exp(i\delta)}{\sqrt{( V_{ud}^{2} + V_{us}^{2})}\sqrt{(1- V_{ub}^{2})}}, \quad V_{cb}\to V_{cb}\nonumber\\
&& V_{td}=\frac{ V_{us}V_{cb}}{\sqrt{( V_{ud}^{2} + V_{us}^{2})}\sqrt{(1-
 V_{ub}^{2})}} -\frac{V_{ud}\sqrt{(1-V_{ub}^{2} - V_{cb}^{2})}V_{ub}\exp(i\delta)}{\sqrt{(V_{ud}^{2} + V_{us}^{2})}\sqrt{(1- V_{ub}^{2})}} 
 \nonumber\\
&& V_{ts}=-\frac{V_{ud}V_{cb}}{\sqrt{( V_{ud}^{2} +
 V_{us}^{2})}\sqrt{(1- V_{ub}^{2})}}-\frac{ V_{us}\sqrt{(1-V_{ub}^{2} -
 V_{cb}^{2})}V_{ub}\exp(i\delta)}{\sqrt{( V_{ud}^{2} + V_{us}^{2})}\sqrt{(1- V_{ub}^{2})}} \nonumber\\
  &&V_{tb}= \sqrt{(1-V_{ub}^{2} - V_{cb}^{2})}
\label{CKM}
\eea
 Since $s_{12}^{2}+ c_{12}^{2}=1 $, we have 
\be
V_{ud}^{2} + V_{us}^{2}= c_{13}^{2} = 1- V_{ub}^{2}
\label{Vub}
\ee
We note that this expression could be used to obtain $V_{ub}$ provided 
that  $V_{ud}$ and $V_{us}$ could be  measured with great precision
which could be achieved in  future experiments. With this relation,
we recover the expressions for $V_{ud}$ and $V_{us}$ in Eq. (\ref{CKM}),
given here with  the factor $\sqrt{(1- V_{ub}^{2})}/\sqrt{( V_{ud}^{2} +
V_{us}^{2})}$ included  so that  unitarity for the CKM matrix 
is explicitly satisfied.
\bigskip

 The above expressions allow a direct determination of 
$V_{cd}$, $ V_{cs}$, $ V_{td}$, $V_{ts}$ and $ V_{tb}$ in terms of 
$V_{ud}$, $ V_{us}$, $ V_{ub}$,  $V_{cb}$. In the following we shall
present  expressions and results of our analysis for 
$|V_{td}/V_{ts}|$, $\sin\beta$, $\sin\alpha$ and  
$\sin\gamma$ in terms of the known  CKM matrix 
elements provided by the global fit and the CP-violating phase $\delta$.

 As seen from the above expressions, the CP-violating phase $\delta$
originating from $V_{ub}$ enters in the quantities $V_{td}$, $V_{ts}$
and also in $V_{cd}$, $V_{cs}$ and manifests itself in  
the  CP-asymmetries  in $B^{0}-\bar{B}^{0}$
mixing and in charmless $B$ decays, for example. 

Consider now the quantity $|V_{td}/V_{ts}|$. We have from Eq. (\ref{CKM}):
\be
|V_{td}/V_{ts}| =\frac{V_{us}}{V_{ud}}\biggl(\frac{\sqrt{(1-2\,K_{d}\cos\delta+K_{d}^2)} }{\sqrt{(1+2\,K_{s}\cos\delta+K_{s}^2)}}\biggr)
\label{Rtd}
\ee
with
\be
K_{d} =
\frac{V_{ud}V_{ub}}{V_{us}V_{cb}}\sqrt{(1-V_{ub}^{2}-V_{cb}^{2})}, 
 \quad  K_{s} =
\frac{V_{us}V_{ub}}{V_{ud}V_{cb}}\sqrt{(1-V_{ub}^{2}-V_{cb}^{2})},
\label{Kd}
\ee
For the $(db)$ unitarity triangle, the angles $\alpha,\beta,\gamma$
 are given by \cite{PDG,Nir}
\be
 \alpha =\arg(-V_{td}V_{tb}^{*}/V_{ud}V_{ub}^{*}), \quad 
 \beta =\arg(-V_{cd}V_{cb}^{*}/V_{td}V_{tb}^{*}), \quad
 \gamma =\arg(-V_{ud}V_{ub}^{*}/V_{cd}V_{cb}^{*})
\label{db}
\ee
with the sides:
\be
 R_{u}= |V_{ud}V_{ub}^{*}/V_{cd}V_{cb}^{*}|, \quad R_{t}=
|V_{td}V_{tb}^{*}/V_{cd}V_{cb}^{*}|, \quad  R_{c}= 1
\label{Ru}
\ee
Using the expressions in Eq. (\ref{CKM}), we find:
\bea
&& \sin\alpha =
\frac{\sin\delta}{\sqrt{(1-2\,K_{t}\cos\delta + K_{t}^2)}}, \quad
\sin\gamma =\frac{\sin\delta}{\sqrt{(1 + 2\,K_{c}\cos\delta + K_{c}^2)}}\nonumber\\
&& \sin\beta = \frac{(K_{t} +  K_{c})\sin\delta}{\sqrt{(1-2\,K_{t}\cos\delta +
  K_{t}^2)}\sqrt{(1+2\,K_{c}\cos\delta + K_{c}^2)}} \
\label{sin}
\eea
with
\be
K_{t} =
\frac{V_{ud}V_{ub}}{V_{us}V_{cb}}\sqrt{(1-V_{ub}^{2}-V_{cb}^{2})}, 
 \quad  K_{c} =
\frac{V_{ud}V_{ub}V_{cb}}{V_{us}\sqrt{(1-V_{ub}^{2}-V_{cb}^{2})}},
\label{Ktc}
\ee
($K_{t}+ K_{c} = V_{ud}V_{ub}(1-V_{ub}^{2})/(V_{us}V_{cb}
\sqrt{(1-V_{ub}^{2}-V_{cb}^{2})}) $)

Similarly, we have:
\bea
&& R_{u}=\frac{(K_{c}+K_{t})}{\sqrt{(1+2\,K_{c}\cos\delta  + K_{c}^2)}},\nonumber \\
&&  R_{t}= \frac{\sqrt{(1-2\,K_{t}\cos\delta +  K_{t}^2)}}{\sqrt{(1+2\,K_{c}\cos\delta + K_{c}^2)}}, \quad  R_{c}= 1.
\label{sides}
\eea
From the expressions for $\sin\alpha$, $\sin\beta$ and $\sin\gamma$ 
in Eq. (\ref{sin}), we recover the relation 
\be
 \sin\beta= R_{u}\sin\alpha, \quad \sin\gamma= R_{t}\sin\alpha
\label{beta}
\ee
obtained previously in \cite{Buras}.

 Eq. (\ref{sin}) for $\sin\gamma$ can be used to express $\delta$
in terms of $\gamma$. We have:
\be
\cos\delta = \cos\gamma(\sqrt{(1 -K_{c}^{2}\sin^{2}\gamma)})  
-K_{c}\sin^{2}\gamma
\label{delta}
\ee
 
 As can be seen from Eq. (\ref{Ktc}), $K_{c}$ is
very much suppressed compared with $K_{t}$ ($K_{c} = 0.006$, $K_{t} = 0.366$), 
 $\sin\delta\approx \sin\gamma$ and $\cos\delta\approx \cos\gamma$
with  $\sin^{2}\gamma$ and $\cos\gamma$ correction terms of
$O(10^{-4})$. Hence  $\delta$ can be replaced by $\gamma$ in 
 the computed quantities with no great  loss of accuracy. Thus, with 
$\sin\delta=  \sin\gamma$ and $\cos\delta= \cos\gamma$, we
obtain, with the global fit central value $V_{ub}=0.00347$:
\bea
&& V_{cd} = -0.225112 - 0.000138\,\exp(i\gamma), \quad 
V_{cs}= 0.973467 - 0.000032\,\exp(i\gamma) \label{Vcd}\\
&&V_{td} = 0.009237 - 0.003378\,\exp(i\gamma), \quad 
V_{ts}=  -0.039946 - 0.000781\,\exp(i\gamma) \label{Vtd}
\eea
We see that the contribution from the suppressed CKM matrix elements
$V_{cb}$ and $V_{ub}$ produces only a small contribution to 
$V_{cd}$ and $V_{cs}$. Thus unitarity of the CKM matrix in the standard model
with 3 generations implies that $V_{cd}$ and $V_{cs}$  are rather well
determined and therefore could be used in the
semi-leptonic and non-leptonic decays of charmed mesons and charmed 
baryons.\break In particular the semi-leptonic $D_{s}$ and $D$ decays 
can be used to obtain the decays constants $f_{D}$ and $f_{D_{s}}$.

\begin{figure}[ht]
\centering
\includegraphics[height=8.0cm,angle=0]{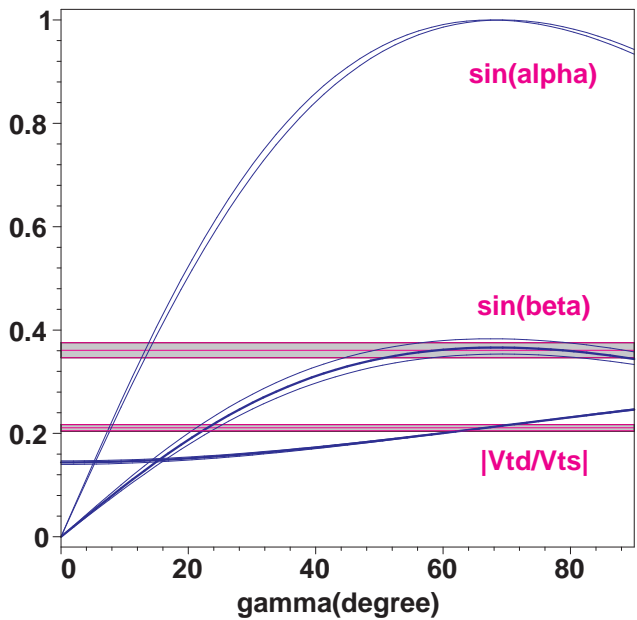}
\caption{$|V_{td}/V_{ts}|$, $\sin\beta$ and $\sin\alpha$ plotted against
the angle $\gamma$.  The middle curves
represent the computed  quantities for $V_{ub}=0.00347$, the global
fit central value; the upper and lower curves for 
for $V_{ub}=0.00363$ and  $V_{ub}= 0.00335$ respectively. The 
deviation from the central value for $|V_{td}/V_{ts}|$ and 
$\sin\alpha$ is barely visible. The  solid straight lines are the 
measured values with experimental errors represented by the gray area.}
\label{Fig2}
\end{figure}

 The matrix element $V_{td}$, gets a large imaginary part from 
the phase $\delta$ in $V_{ub}$ as given in Eq. (\ref{Vtd}). This large
imaginary part is  responsible for the large time-dependent CP-asymmetry
in $B$ decays. For the $|V_{td}/V_{ts}|$,   Eq. (\ref{Rtd}) gives, with 
$\delta$ replaced by $\gamma$:
\be
|V_{td}/V_{ts}|
=\frac{V_{us}}{V_{ud}}\biggl(\frac{\sqrt{(1-2\,K_{d}\cos\gamma + K_{d}^2)}
  }{\sqrt{(1+ 2\,K_{s}\cos\gamma + K_{s}^2)}}\biggr)
\label{RVt}
\ee
and by neglecting the suppressed  $K_{c}$ terms, from Eq. (\ref{beta}):
\bea
&& \sin\beta=
\frac{B_{u}\sin\gamma}{\sqrt{(1-2\,K_{t}\cos\gamma + K_{t}^2)}}, \quad
B_{u} =\frac{V_{ud}V_{ub}(1-V_{ub}^{2})}{V_{us}V_{cb}\sqrt{(1-V_{ub}^{2}-V_{cb}^{2})}}\nonumber\\
&& \sin\alpha= \frac{\sin\gamma}{\sqrt{(1-2\,K_{t}\cos\gamma + K_{t}^2)}}.
\label{alpha}
\eea
It is clear from Eq. (\ref{alpha}) that $\sin\beta$ is  a 
measure of  $\gamma$,  $V_{ub}$ and  $V_{ub}/V_{cb}$. 
Numerically, $B_{u}= 0.366$,
  $\sin\beta\approx 0.366\,\sin\alpha$ as seen from the  plot in 
Fig. (\ref{Fig2}) . On the other hand, the curve for $|V_{td}/V_{ts}|$ 
 in the plot implies that $\gamma=(68.6\pm 3)^{\circ}$. The middle 
curve for $\sin\beta$ shows  the measured $\sin\beta$ at almost 
the same value of $\gamma$. The lower curve for $\sin\beta$ is  a bit 
below the measured value but consistent with   
experiment. The  upper curve is slightly  above the measured
value. This indicates that a value for $V_{ub}$ not too far from 
the global fit is favored.

In conclusion, by expressing the CKM matrix elements  in terms of 
$V_{ud}$, $ V_{us}$, $ V_{ub}$, $V_{cb}$ and the CP-violating 
phase  $\delta $ instead of using the usual 
mixing angle parametrization, we have obtained  analytical expressions 
for the CKM matrix elements and the angles of the $(db)$ unitarity 
angle. Using  the global fit for $V_{ud}$, $ V_{us}$, $ V_{ub}$,
$V_{cb}$  and the measured values for $|V_{td}/V_{ts}|$ and $\sin\beta$, 
we find  $\gamma=(68.6\pm 3)^{\circ}$ at which $\sin\beta$ and $\sin\alpha$
are near its maximum. Our simple and direct approach  could be 
used in any further analysis of the CKM matrix elements with 
 more precise measured values for $V_{ub}$ and  $V_{cb}$.


\end{document}